%% file: main.tex
\documentclass[reprint,preprintnumbers,nofootinbib,amsmath,amssymb,aps,prd,]{revtex4-2}
\usepackage[T1]{fontenc} 
\usepackage{hyperref}
\usepackage{epsfig}
\usepackage{graphicx}
\usepackage{amsmath}
\usepackage{amssymb}
\usepackage{algorithm}
\usepackage{algorithmic}
\usepackage{bbm}
\usepackage{amsfonts}
\usepackage{mathrsfs}
\usepackage{multirow}
\usepackage{booktabs}
\usepackage{siunitx}
\usepackage{newtxmath}
\usepackage{xcolor}
\usepackage{soul}
\usepackage{lipsum}
\usepackage{graphicx}
\graphicspath{{./}{figures/}}

\begin{document}

\title{Inference of maximum parsimony phylogenetic trees with model-based classical and quantum methods}

\author{Jiawei Zhang{$^{1,2\dag}$}\thanks{These authors contributed equally to this work.}}
\author{Yibo Chen{$^{2\dag}$}}
\author{Yang Zhou{$^{2}$}}
\author{Jun-Han Huang{$^{2}$}\thanks{Corresponding author}}
\email[]{huangjunhan@genomics.cn}

\affiliation{
    {$^{1}$}University of Chinese Academy of Sciences, Beijing 101408, China \\
    {$^{2}$}State Key Laboratory of Genome and Multi-omics Technologies, BGI Research, Shenzhen 518083, China
}

\date{\today}

\begin{abstract}
The maximum parsimony phylogenetic tree reconstruction problem is NP-hard, presenting a computational bottleneck for classical computing and motivating the exploration of emerging paradigms like quantum computing. To this end, we design three optimization models compatible with both classical and quantum solvers. Our method directly searches the complete solution space of all possible tree topologies and ancestral states, thereby avoiding the potential biases associated with pre-constructing candidate internal nodes. Among these models, the branch-based model drastically reduces the number of variables and explicit constraints through a specific variable definition, providing a novel modeling approach effective not only for phylogenetic tree building but also for other tree problems. The correctness of this model is validated with a classical solver, which obtains solutions that are generally better than those from heuristics on the $\textit{GAPDH}$ gene dataset. Moreover, our quantum simulations successfully find the exact optimal solutions for small-scale instances with rapid convergence, highlighting the potential of quantum computing to offer a new avenue for solving these intractable problems in evolutionary biology.

\end{abstract}

\maketitle
\input{Introduction}

\input{Results}

\input{Discussion}

\input{Methods}

\section*{Data availability}
The $\textit{GAPDH}$ gene sequence data used in this paper can be obtained from the NCBI database. Direct URL to data: \url{https://www.ncbi.nlm.nih.gov/gene/2597/ortholog/?scope=7742}.

\section*{Code availability}
The entire code package for this paper is available in the GitHub repository: \url{https://github.com/DemonCass/Phylogenetic-tree}.

\begin{acknowledgments}
The authors would like to thank Man-Hong Yung, Xian-Zhe Tao and Qinyuan Zheng for helpful discussions.
\end{acknowledgments}

\section*{AUTHOR CONTRIBUTIONS}
\noindent Conceptualization, J.Z., Y.C. and J.-H.H.; methodology: J.Z., Y.C. and J.-H.H.; software: J.Z.; validation, J.Z., Y.C. and J.-H.H.; investigation, J.Z., Y.C. and J.-H.H.; resources, Y.C., Y.Z. and J.-H.H.; data curation, J.Z.; writing – original draft, J.Z.; writing – review \& editing, J.Z., Y.C., Y.Z. and J.-H.H.; visualization, J.Z. and J.-H.H.; supervision: J.-H.H.; project administration, J.-H.H..

\section*{Competing interests}
\noindent The authors declare no competing interests.

\bibliography{egbib}

\end{document}

%% file: Introduction.tex
\section{Introduction}
Phylogenetic tree reconstruction, the inference of evolutionary relationships, is a cornerstone of modern biology with profound implications in fields such as species identification, disease tracking, biodiversity conservation and drug discovery \cite{felsenstein2004inferring,bull2001applied,davies2008phylogenetic,o2020corbomycin,saslis2011use}. Among the various reconstruction methods, maximum parsimony remains a primary approach due to its intuitive logic, its independence from the explicit evolutionary models required by methods like maximum likelihood or bayesian inference, and its robust performance when evolutionary changes are rare and homoplasy is minimal \cite{cancino2010multi,puttick2019probabilistic}.

Despite the conceptual advantages of maximum parsimony, finding the most parsimonious tree is an NP-hard problem \cite{day1986computational}, creating a significant computational bottleneck for classical computing. While heuristic algorithms are commonly used to handle this complexity \cite{nei2000molecular}, their effectiveness diminishes in datasets with a large number of species because the attraction basin for each optimum shrinks dramatically, making the best solutions increasingly difficult to find \cite{kirkup2000rolling}. This limitation drives the search for novel computational paradigms designed to locate optimal or high-quality solutions with greater efficiency.

One promising avenue for addressing these computational bottlenecks is quantum computing, which is increasingly recognized for its potential applications in genomics and life sciences \cite{emani2021quantum, fedorov2021towards, marchetti2022quantum, nalkecz2024quantum, damborsky2025quantum, bose2026advancing, maurizio2025quantum}. By leveraging superposition and entanglement, a quantum processor with $N$ qubits operates within a state space of dimension $2^N$. Crucially, computational speedups derive not merely from parallel exploration, but also from the ability to utilize quantum interference to enhance the probability amplitude of the optimal solution \cite{shor1999polynomial, nielsen2010quantum, montanaro2016quantum, rastegin2017role, dalzell2023quantum}. Within this framework, hybrid quantum-classical algorithms have been extensively investigated for solving complex combinatorial optimization problems. While challenges remain, these approaches offer a feasible pathway for exploring quantum utility on noisy intermediate-scale quantum (NISQ) devices \cite{preskill2018quantum, bharti2022noisy, gemeinhardt2023quantum, abbas2024challenges, blekos2024review}.

Solving the maximum parsimony problem with quantum algorithms first requires an efficient mathematical model, as its complexity fundamentally dictates the performance of solver. We note that some previous studies map this problem to the graph-theoretic Steiner tree problem. However, the Steiner tree problem in graphs is a classical NP-hard problem \cite{foulds1982steiner, hwang1992steiner}, and a core challenge lies in handling the potential ancestral nodes (Steiner points). The common strategy of pre-constructing a finite set of candidate ancestral nodes is flawed. If the true optimal ancestral sequence is not in the pre-defined set, the resulting MP tree is not guaranteed to be optimal. This pre-processing step is not only costly but also introduces bias \cite{catanzaro2013mixed, bach2024quantum}. Furthermore, these models involve numerous constraints and do not infer ancestral sequences during the search process.

\section{Results}
\begin{figure*}[t!]
    \centering
        \includegraphics[height=14.5cm, keepaspectratio]{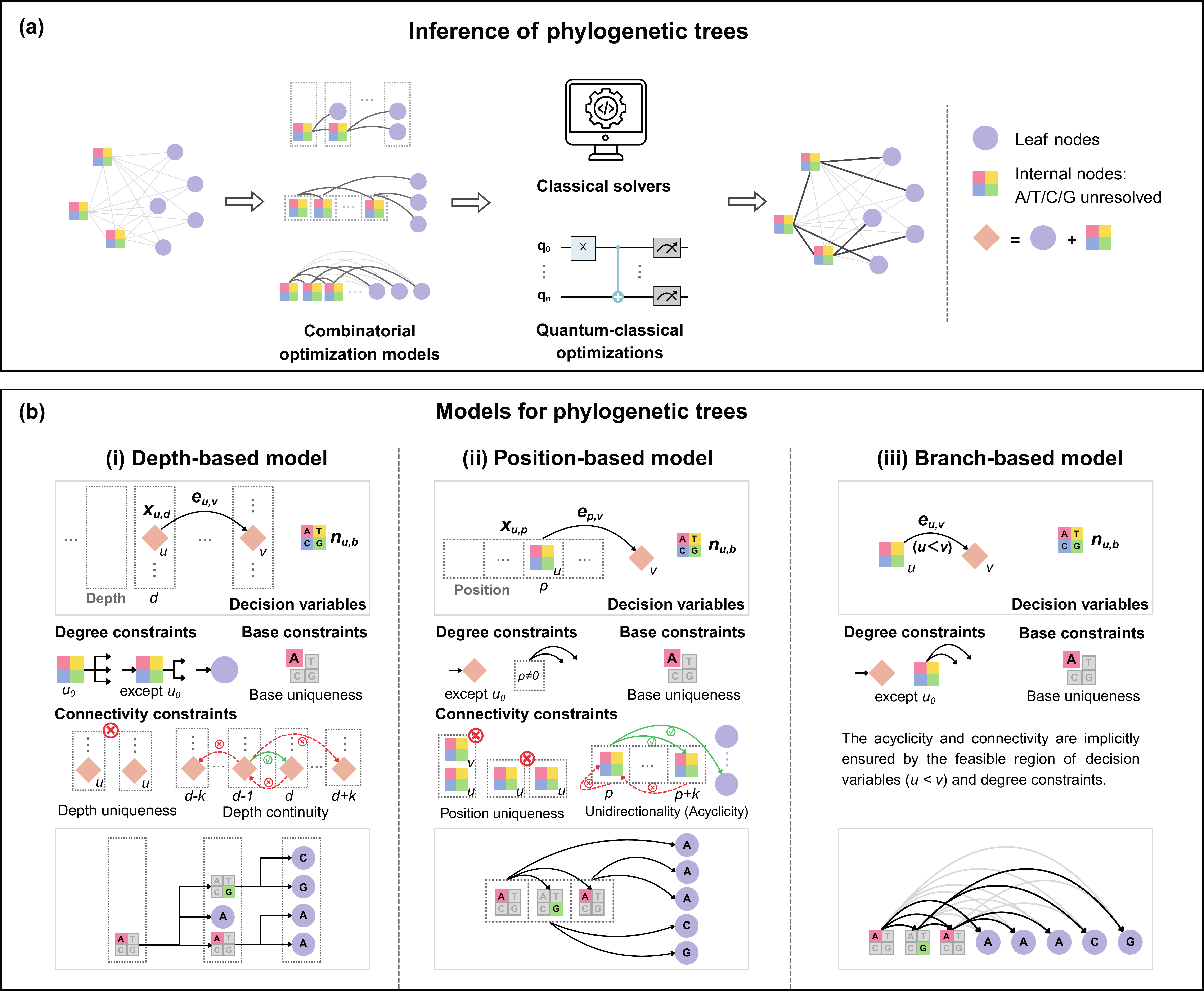}
        \caption{Schematic of the model-based framework for maximum parsimony phylogenetic trees inference. (a) The problem is converted into a combinatorial optimization model, which can use classical or quantum optimizers to find the globally maximum parsimony phylogenetic tree. (b) Detailed comparison of the three models designed for this problem: (i) the depth-based model arranges nodes by depth, with each depth $d\ge1$ containing at most $3\cdot2^{d-1}$ nodes; (ii) the position-based model assigns a unique position to each internal node; (iii) the branch-based model directly defines the connections. The branch-based model is the most efficient, as it implicitly ensures a valid tree structure with fewer constraints.}
    \label{fig:Schematic_diagram}
\end{figure*}

To overcome these hurdles, we propose three optimization models that simultaneously infer ancestral sequences while constructing the tree topology: the depth-based, position-based and branch-based models, as illustrated in Fig. \ref{fig:Schematic_diagram}. As the branch-based model is particularly efficient, we validate it using a classical solver against the branch-and-bound algorithm \cite{hendy1982branch} and heuristics to confirm its correctness and assess its performance. Furthermore, we explore the feasibility of a quantum pathway by implementing the model with variational quantum algorithms. This study aims to investigate whether these quantum approaches can offer a novel and effective method for solving intractable phylogenetic problems.

%% file: Results.tex
\subsection{Model}
We formulate the reconstruction of a phylogenetic tree as a combinatorial optimization problem under the maximum parsimony criterion. The objective is to minimize the parsimony score subject to the constraints that define a valid phylogenetic tree topology.

A phylogenetic trees can be either rooted or unrooted. In these trees, the leaf nodes typically represent extant species, while the internal nodes represent their extinct or hypothetical ancestors. These nodes are connected by edges symbolizing evolutionary lineages, and the transformations occurring along these edges are the substitutions that the parsimony score quantifies.

Our model is based on the properties of an unrooted phylogenetic tree, it is composed of a set of $n$ leaf nodes $L$ and $n-2$ internal nodes $I$ for a total of $|V|=2n-2$ nodes \cite{kannan2012maximum}. For any given position in the sequence alignment, the character state at each node is selected from a set $B=\{A,C,G,T,-\}$, where $'-'$ represents an indel or an ambiguous nucleotide. 

\begin{table}[htbp]
    \centering
        \begin{tabular}{|c|c|*{5}{>{\centering\arraybackslash}m{0.9cm}|}}
            \hline
            \multicolumn{2}{|c|}{} & \multicolumn{5}{c|}{Node \textit{j}} \\
            \cline{3-7}
            \multicolumn{2}{|c|}{} & A & C & G & T & -\\
            \hline
            \multirow{5}{*}{\rotatebox[origin=c]{90}{Node \textit{i}}} & A & 0 & 2 & 1 & 2 & 4\\
            \cline{2-7}
            & C & 2 & 0 & 2 & 1 & 4\\
            \cline{2-7}
            & G & 1 & 2 & 0 & 2 & 4\\
            \cline{2-7}
            & T & 2 & 1 & 2 & 0 & 4\\
            \cline{2-7}
            & - & 4 & 4 & 4 & 4 & 0\\
            \hline
        \end{tabular}
    \caption{An example of a step matrix $S$ for parsimony analysis. The cost for a substitution from state $i$ to state $j$ is given by $S_{ij}$. This matrix is a prior assumption and can be modified to reflect different evolutionary models or to analyze other sequence types.}
    \label{tab:step_matrix}
\end{table}

In biological reality, different types of substitutions occur with different frequencies \cite{brown1982mitochondrial,wakeley1994substitution}. Therefore, we employ a step matrix $S$ (Table \ref{tab:step_matrix}) to define the cost of changing from one state to another. 

To effectively model the unrooted topology, we orient the tree by selecting an internal node $u_0\in I$ to act as a reference node. Based on this reference, a unique depth and position can be assigned to every other node. Since the resulting directed edges are purely a computational artifact and not representative of the actual evolutionary path, the choice of any internal node as the reference does not alter the final unrooted topology.

\subsubsection{Depth-based model}
The primary challenge in modeling an unrooted tree is to impose a coherent structure that prevents cycles. A common and intuitive strategy is to establish a hierarchy by defining the depth of each node relative to the reference node, and to use constraints to ensure all connections flow in a single direction, thereby obtaining an acyclic structure.

We first establish the hierarchy by fixing the reference node $u_0$ at depth 0. For other node $u\in V\setminus \{u_0\}$, we introduce binary variables $x_{u,d}$ to determine its position. The following constraint then ensures that each of these non-reference nodes is assigned to exactly one depth level $d \in \{1, \dots, n-2\}$:
\begin{equation}
    \sum_{d=1}^{n-2} x_{u, d} = 1, \quad \forall u \in V \setminus \{u_0\}.
    \tag{1} \label{eq:assign_depth}
\end{equation}

To define the connections within the oriented tree, we introduce  binary variables $e_{u,v}$ for each pair of nodes \((u,v)\in V \times V\). A valid tree topology is enforced by the following set of degree constraints:
\begin{equation}
    \tag{2}\label{eq:depth degree}
    \begin{aligned}
        \sum_{u\in V}e_{u, u_0}&= 0,
        \\
        \sum_{v \in V}e_{u_0, v}&= 3,
        \\
        \sum_{ u \in V} e_{u, v}&= 1,\quad \forall v \in V \setminus \{u_0\},
        \\
        \sum_{v \in V} e_{u, v}&= 2,\quad \forall u \in I \setminus \{u_0\},
        \\
        \sum_{v\in V}e_{u, v}&= 0,\quad \forall u\in L.
    \end{aligned}
\end{equation}
These constraints define the in-degree and out-degree for each type of node: the reference node has an in-degree of 0 and an out-degree of 3; other internal nodes have an in-degree of 1 and an out-degree of 2; and leaf nodes have an in-degree of 1 and an out-degree of 0.

To link depth assignments to the tree structure and prevent cycles, any connected pair ($u,v$) must observe the following constraint:
\begin{equation}
    e_{u,v}=1\Longrightarrow\sum_{d=1}^{n-2}(x_{u,d-1}\cdot x_{v,d})=1,\quad\forall u,v\in V.
    \tag{3}
\end{equation}

Finally, let binary variables $n_{u,b}$ indicate that internal node $u$ is assigned base $b$. Each internal node must be assigned exactly one base:
\begin{equation}
    \sum_{b\in B}n_{u, b}=1, \quad \forall u\in I.
    \tag{4} \label{eq:base}
\end{equation}

With the variables for the tree structure and base assignments defined, the objective of the model is to minimize the total parsimony score ($H_{p_1}$). This score is calculated based on the step matrix $S$ as follows:
\begin{equation}
    \begin{aligned}
        H_{p_1} =& \sum_{u \in I} \sum_{v \in L} \sum_{b\in B} S_{g(v)b} e_{u, v} n_{u, b}\
        \\
       +& \sum_{u \in I} \sum_{v \in I} \sum_{b\in B}\sum_{b'\in B} S_{bb'} e_{u, v} n_{u, b} n_{v, b'},
    \end{aligned}
    \tag{5}
    \label{eq:H_pars_1}
\end{equation}
where $g(v)$ is the given base of leaf node $v$. This score sums the substitution costs over all edges, distinguishing between edges connecting to leaves and those between two internal nodes.

The previously defined topological and assignment constraints (Eq. \ref{eq:assign_depth} - Eq. \ref{eq:base}) are incorporated as quadratic penalty terms. The complete depth-based model is:
\begin{equation}
\tag{6}
    \begin{aligned}
        H_1 &= H_{p_1} + P \Biggl\{ \sum_{u \in V \setminus \{u_0\}} ( 1 - \sum_{d=1}^{n-2} x_{u, d} )^2 + \sum_{u\in V}e_{u,u_0}
        \\
        &+(3-\sum_{v\in V}e_{u_0,v})^2 +\sum_{v\in V \setminus \{u_0\}}(1-\sum_{u\in V} e_{u,v})^2
        \\
        &+\sum_{u\in I \setminus \{u_0\}}(2-\sum_{v\in V}e_{u,v})^2 + \sum_{u\in L}\sum_{v\in V}e_{u,v}
        \\
        &+\sum_{u\in V}\sum_{v\in V}e_{u,v}(1-\sum_{d=1}^{n-2}x_{u,d-1}x_{v,d})^2 + \sum_{u \in I} ( 1 - \sum_{b\in B} n_{u, b} )^2 \Biggr\},
    \end{aligned}
\end{equation}
where $P$ is a penalty factor.

The primary drawback of the depth-based model is its significant computational inefficiency. This arises from the excessive number of variables and penalty terms required, which scale rapidly with the number of species.

\subsubsection{Position-based model}
As an alternative to the depth-based method, we can assign positions to the nodes. Since leaf nodes have only a single incoming edge, their positions do not need to be assigned as variables, which significantly reduces the total number of variables required.

In the position-based model, we assign each non-reference internal node to a unique position $p\in \{1,\dots,n-3\}$ using binary variables $x_{u,p}$. This creates a bijective mapping between nodes and positions, while  the reference node $u_0$ is fixed at position 0.
\begin{equation}
    \begin{aligned}
        \sum_{u\in I\setminus\{u_0\}}x_{u, p} &= 1, \quad\forall p\in \{1,\dots,n-3\},
        \\
        \sum_{p=1}^{n-3}x_{u, p} &= 1, \quad \forall u\in I\setminus\{u_0\}.
    \end{aligned}
    \tag{7} \label{eq:position}
\end{equation}

Next, the connectivity of the tree is defined using the binary variable $e_{p,u}$, which represents a directed edge from a position $p$ to a node $u\in V\setminus\{u_0\}$. To ensure that these edges form a valid tree structure, we impose the following degree constraints:
\begin{equation}
    \begin{aligned}
        \sum_{p=0}^{n-3}e_{p, u} &= 1,\quad \forall u\in V\setminus \{u_0\},
        \\
        \sum_{u\in V \setminus \{u_0\}}e_{p, u} &= 2, \quad \forall p\in \{1,\dots,n-3\}.
    \end{aligned}
    \tag{8} \label{eq:position degree}
\end{equation}

To prevent cycles, an edge $e_{p,u}$ is permitted only if the position index of node $u$ is greater than $p$:
\begin{equation}
    e_{p,u}=1\Longrightarrow\sum_{p'=0}^{p}x_{u, p'} = 0,\quad \forall u\in I,\forall p \in \{0,\dots,n-3\}.
    \tag{9} \label{eq:position cycle}
\end{equation}

The objective function for the position-based model is the parsimony score, $H_{p_2}$:
\begin{equation}
    \begin{aligned}
        H_{p_2} =& \sum_{u \in I} \sum_{v \in L} \sum_{p=0}^{n-3}\sum_{b\in B} S_{g(v)b} x_{u,p}e_{p, v} n_{u, b}\
        \\
        + & \sum_{u \in I} \sum_{v \in I} \sum_{p=0}^{n-3} \sum_{b\in B}\sum_{b'\in B} S_{bb'} x_{u,p} e_{p, v} n_{u, b} n_{v, b'}.
    \end{aligned}
    \tag{10}
    \label{eq:H_pars_2}
\end{equation}

\begin{figure*}[t!]
    \centering
    \includegraphics[height=4cm, keepaspectratio]{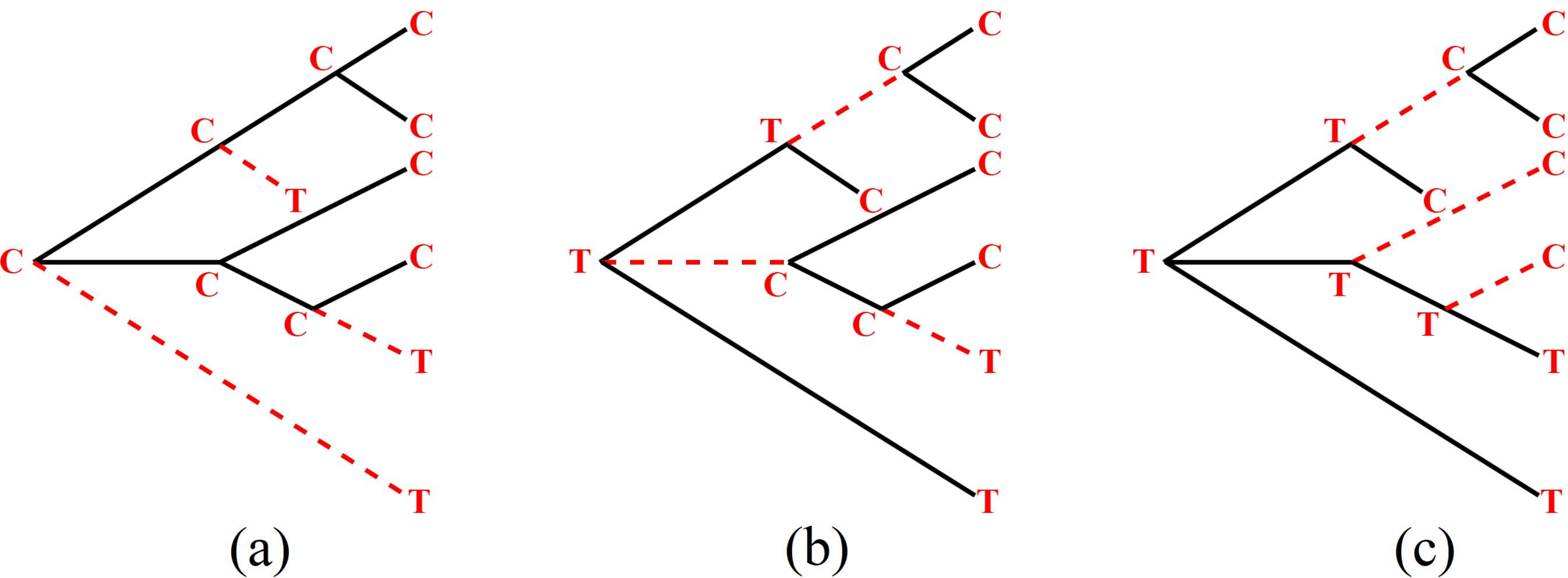}
    \vspace{5pt}
    \caption{Multiple equivalent optimal solutions may exist at a single site. After reconstruction of a single site, although the bases of the internal nodes in (a) (b) and (c) are not identical, they all have the same minimum number of substitutions. The red dashed lines indicate the branches that underwent evolutionary changes.}
    \label{fig:optimal_solutions}
\end{figure*}

Finally, we integrate $H_{p_2}$ with penalty terms for all constraints to formulate the complete position-based model:
\begin{equation}
\tag{11}
\begin{aligned}
    H_2 &= H_{p_2} + P \Biggl\{ \sum_{p=1}^{n-3} ( 1 - \sum_{u\in I \setminus \{u_0\}} x_{u, p} )^2 + \sum_{u\in I \setminus \{u_0\}} ( 1 - \sum_{p=1}^{n-3}x_{u, p} )^2
    \\
    &+ \sum_{u\in V\setminus\{u_0\}}(1-\sum_{p=0}^{n-3}e_{p,u})^2 + \sum_{p=1}^{n-3}(2-\sum_{u\in V\setminus\{u_0\}}e_{p,u})^2
    \\
    & + \sum_{p=0}^{n-3}\sum_{u\in I}e_{p,u}\sum_{p'=0}^{p}x_{u,p} + \sum_{u \in I} ( 1 - \sum_{b\in B} n_{u, b} )^2 \Biggr\}.
\end{aligned}
\end{equation}

Although this model simplifies some of the constraints, its objective function contains more higher-order interactions, which increases the computational difficulty of solving the problem.

\subsubsection{Branch-based model}
The depth-based model requires too many variables and constraints, while the objective function of the position-based model is overly complex. Observing these challenges, we further propose a highly simplified branch-based model.

Since the internal nodes are essentially identical before the base or sequence information is determined, we can establish a unique integer index for each node and define binary variables $e_{u,v}$ to represent a direct edge between internal node $u$ and non-reference node $v$, with the crucial condition that the index of $v$ must be greater than the index of $u$ ($v>u$).

This variable definition naturally includes the following constraints:
\begin{itemize}
    \item Implicit acyclicity: The condition $v>u$ inherently prevents cycles, because any path through the tree must follow a sequence of nodes with strictly increasing indices.
    \item Implicit degree constraints: Since edges can only originate from internal nodes, the out-degree of all leaf nodes is guaranteed to be 0. Similarly, edges only connect to non-reference nodes, the in-degree of the reference node is guaranteed to be 0.
\end{itemize}
This design significantly simplifies the complexity of the model by eliminating the need for the explicit acyclicity and connectivity constraints seen in the previous models.

Notably, the out-degree of the reference node does not require an explicit constraint, as it's automatically determined by the degrees of all other nodes:
\begin{align*}
        \text{Out}(u_0) &= \underbrace{\sum_{v \in V \setminus \{u_0\}} \text{In}(v)}_{\text{Total in-degrees of others}} - \underbrace{\sum_{u \in I \setminus \{u_0\}} \text{Out}(u)}_{\text{Total out-degrees of others}}
        \\
        &= (2n-3) - 2(n-3) = 3.
\end{align*}

As a result, the correct tree topology can be enforced with just two constraints:
\begin{equation}
\tag{12}
    \begin{aligned}
        \sum_{u \in I} e_{u, v} &= 1, \quad \forall v \in V \setminus \{u_0\},
        \\
        \sum_{v \in V \setminus \{u_0\}} e_{u, v} &= 2, \quad \forall u \in I \setminus \{u_0\}.
    \end{aligned}
\end{equation}

The objective function and the complete branch-based model have been greatly simplified:
\begin{equation}
    \begin{aligned}
       H_{p_3} =& \sum_{u \in I} \sum_{v \in L}\sum_{b\in B} S_{g(v)b} e_{u, v} n_{u, b}\
        \\
       +& \sum_{u \in I} \sum_{v \in I} \sum_{b\in B} \sum_{b'\in B} S_{bb'} e_{u, v} n_{u, b} n_{v, b'},
    \end{aligned}
    \tag{13}
    \label{eq:13}
\end{equation}
\begin{equation}
    \begin{aligned}
    H_3 &= H_{p_3} + P \Biggl\{ \sum_{v \in V \setminus \{u_0\}} (1 - \sum_{u \in I} e_{u, v})^2 
    \\
    & + \sum_{u \in I \setminus \{u_0\}} (2 - \sum_{v \in V \setminus \{u_0\}} e_{u, v})^2+ \sum_{u \in I} ( 1 - \sum_{b\in B} n_{u, b} )^2 \Biggr\},
    \end{aligned}
    \tag{14}
\end{equation}
where the objective function defined in Eq. (\ref{eq:13}) contains high-order terms.

\begin{table}[htbp]
    \centering
        \small
        \renewcommand{\arraystretch}{1.2} 
        \begin{tabular}{c|ccc}
            \hline
            \multirow{2}{*}{\textbf{\begin{tabular}[c]{@{}c@{}}Leaf nodes\end{tabular}}} & \multicolumn{3}{c}{\textbf{\# of variables}} \\
            \cline{2-4}
            & \textbf{Depth-based} & \textbf{Position-based} & \textbf{Branch-based} \\
            \hline
            50 & 14,500 & 7,105 & 3,768 \\
            100 & 59,000 & 29,205 & 15,043 \\
            500 & 1,495,000 & 746,005 & 375,243 \\
            $\cdots$ & $\cdots$ & $\cdots$ & $\cdots$ \\
            $n$ & $6n^2 - 10n$ & $3n^2 - 8n + 5$ & $\frac{3}{2}n^2 + \frac{1}{2}n - 7$ \\
            \hline
        \end{tabular}
        \caption{Comparison of the total number of variables required by each model as the number of leaf nodes varies. The branch-based model offers a significant reduction in the total number of variables compared to the other two models.}
    \label{tab:model_comparison}
\end{table}

As can be intuitively seen from Table \ref{tab:model_comparison}, the branch-based model holds an advantage in its total number of variables. Considering the limited computational resources, the experiments were performed using this model.

\subsection{Model validation}
Having established the branch-based model as our most efficient model, the crucial next step is to validate its correctness. We first test the model by focusing on the single-site maximum parsimony problem.

A key consideration in maximum parsimony is the potential for multiple optimal solutions. For a single site, several different ancestral state reconstructions can yield the same minimal number of substitutions, as illustrated in Fig. \ref{fig:optimal_solutions}. Therefore, our focus is on finding a solution with the minimum total number of mutations rather than on reconstructing a specific set of ancestral states.

\begin{figure}[htbp]
    \centering
        \includegraphics[width=0.8\linewidth]{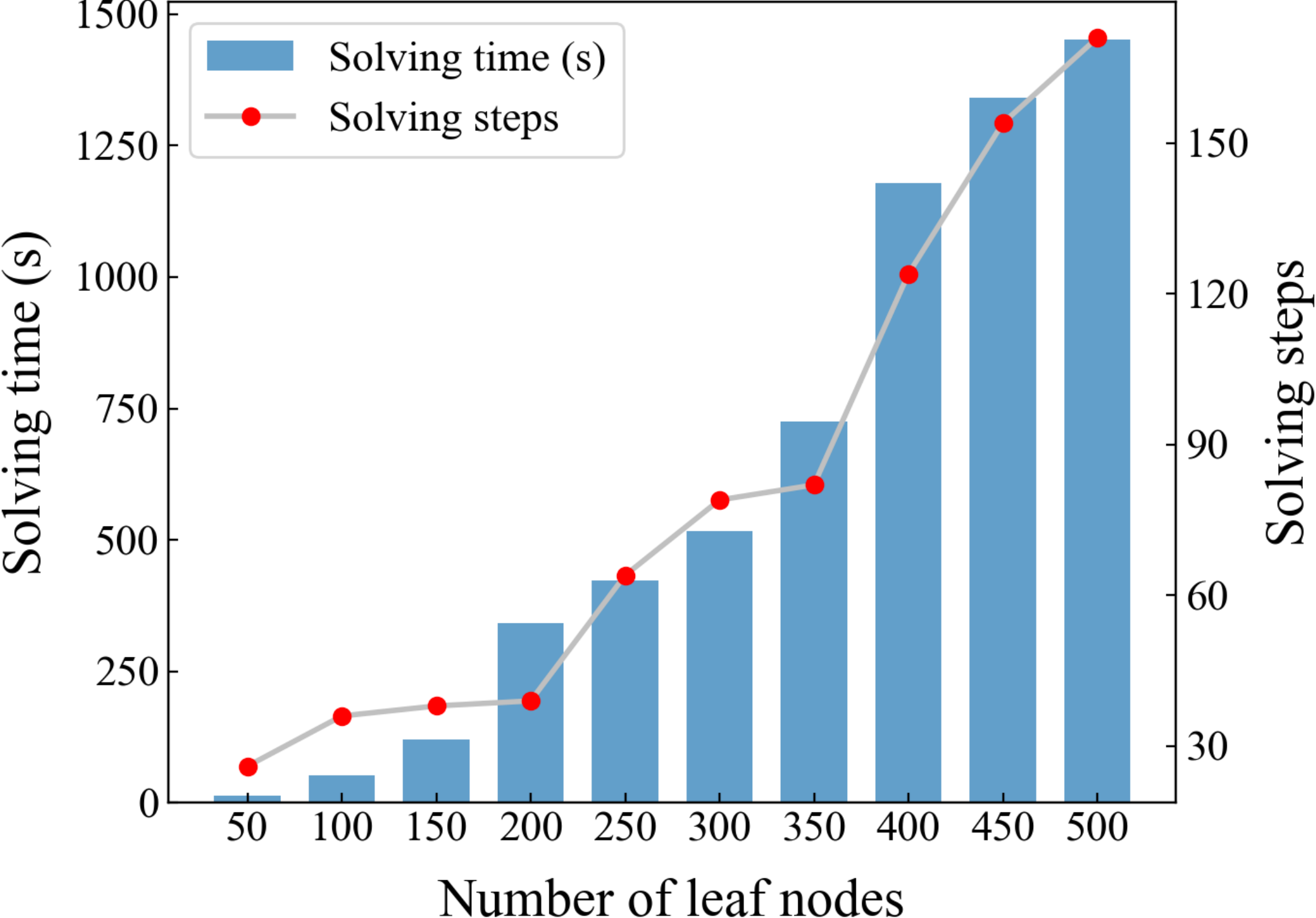}
        \caption{Performance of the classical CP-SAT solver on the branch-based model. The initial base states of the leaf nodes are randomly generated, and each result is the average of ten trials. Each solving step includes methods such as constraint propagation and conflict analysis to reduce the search space.}
    \label{fig:ortools_performance}
\end{figure}

To solve our model, we selected the open-source solver CP-SAT, which is part of the Google OR-Tools optimization suite \cite{cpsatlp}. We benchmark against the guaranteed optimal results from the branch-and-bound algorithm in the MEGA software \cite{tamura2021mega11} and recorded the time and solving steps required for a classical solver to find the optimal solution using our model.

The results presented in Fig. \ref{fig:ortools_performance}. For smaller problem sizes ($n<150$), the solver rapidly identifies the optimal solution. However, as the number of leaf nodes increases, both the solving time and solving steps increase rapidly. A primary reason is that the number of variables and terms in the objective function both grow polynomially with the problem size, as detailed in Table \ref{tab:model_complexity}. The resulting vast search space creates a computational bottleneck, even for a highly optimized classical solver.

\begin{table}[]
    \centering
    \setlength{\tabcolsep}{5pt}
        \begin{tabular}{ccc}
            \toprule
            \textbf{Leaf nodes} & \textbf{\# of variables} & \textbf{\# of terms} \\
            \midrule
            50 & 3,768 & 271,784 \\
            100 & 15,043 & 2,013,459 \\
            500 & 375,243 & 233,866,859 \\
            \dots & \dots & \dots \\
            $n$ & $\mathcal{O}(n^2)$ & $\mathcal{O}(n^3)$ \\
            \bottomrule
        \end{tabular}
        \caption{The relationship between the solving difficulty of the branch-based model and the problem size. The number of variables and terms in the model exhibit polynomial growth as the number of leaf nodes increases.}
    \label{tab:model_complexity}
\end{table} 

The experimental results confirm that our model can successfully identify the maximum parsimony phylogenetic tree for a single site. Furthermore, the challenges in scalability on classical computation underscore the necessity of investigating new computational paradigms.

\subsection{Application to a biological dataset}
Having ensured that an maximum parsimony phylogenetic tree can be obtained for a single site, we now address a more realistic scenario. A tree that is optimal for one site may be suboptimal for another. Therefore, we extend the branch-based model and use $\textit{GAPDH}$ gene sequences from 20 amphibian species, sourced from NCBI.

To extend the model for a sequence fragment of length $m$, we simply expand the base assignment variable from $n_{u,b}$ to the site-specific variable $n_{u, s, b},~s\in\{1,\dots,m\}$. The base uniqueness constraint (Eq. \ref{eq:base}) is then applied to each site $s$ individually, resulting in the final model:
\begin{equation}
    \begin{aligned}
    H_s &= \sum_{u \in I} \sum_{v \in V}\sum_{s=1}^m \sum_{b\in B} \sum_{b'\in B} S_{bb'} e_{u, v} n_{u, s, b} n_{v, s, b'} 
    \\
    & + P \Biggl\{ \sum_{v \in V \setminus \{u_0\}} (1 - \sum_{u \in I} e_{u, v})^2 + \sum_{u \in I \setminus \{u_0\}} (2 - \sum_{v \in V \setminus \{u_0\}} e_{u, v})^2
    \\
    &+ \sum_{u \in I}\sum_{s=1}^m ( 1 - \sum_{b\in B} n_{u, s, b} )^2 \Biggr\},
    \end{aligned}
    \tag{15}
\end{equation}
where the binary variable $n_{u, s, b}$ indicates that internal node $u$ is assigned base $b$ at site $s$. For leaf nodes, this term is not a variable but a pre-defined constant, with $n_{v, s, b'}=1$ if the given sequence data for leaf node $v$ has base $b'$ at site $s$.

The complexity of our model is determined by both the number of species and the sequence length. We therefore segment the multiple sequence alignment into shorter fragments of varying lengths (50-250 bp), using a sliding window approach to generate several distinct datasets for each length. This length range is sufficient to validate the ability of the model to find solutions without being hindered by the known exponential scaling of the problem.

Since the branch-and-bound algorithm is computationally intractable for more than 15 species, we benchmark model against commonly used heuristics, including subtree-pruning-regrafting (SPR), tree-bisection-reconnection (TBR), and Min-mini \cite{nei2000molecular}. While these methods are computationally fast, they do not guarantee finding the globally optimal solution. Therefore, our evaluation directly compares the total number of substitutions found by the different methods.

\begin{table}[]
    \centering
        \renewcommand{\arraystretch}{1.2}
        \begin{tabular}{ccccc}
        \toprule
        \textbf{Fragment} & \multicolumn{4}{c}{\textbf{Average substitutions}} \\
        \cmidrule(lr){2-5}
        \textbf{length} & \textbf{SPR} & \textbf{TBR} & \textbf{Min-Mini} & \textbf{Our model} \\
        \midrule
        50 bp  & 84.4 & 84.4 & 81.8 & \textbf{80.8} \\
        100 bp & 149.0 & 149.2 & 146.0 & \textbf{138.8} \\
        150 bp & 275.6 & 276.4 & 280.8 & \textbf{271.6} \\
        200 bp & 363.6 & 365.0 & 361.2 & \textbf{349.2} \\
        250 bp & 445.6 & 445.6 & 451.2 & \textbf{433.8} \\
        \bottomrule
        \end{tabular}
        \caption{Comparison of the average number of substitutions obtained by different methods. Values are averaged over five independent replicate datasets for each fragment length. The best value for each length is highlighted.}
    \label{tab:multibase_comparison}
\end{table}

Although the true global optimum cannot be determined for these problems due to the limitations of exact algorithms, Table \ref{tab:multibase_comparison} demonstrates that our model consistently finds higher-quality solutions than these common heuristics. 

While this approach improves the solution quality, it does not offer an advantage in solving time. This classical performance trade-off motivates the exploration of alternative computing paradigms. We next attempt to solve this problem using quantum algorithms to explore new pathways and potential computational advantages for this intractable problem.

\subsection{Variational quantum algorithm}
To solve the maximum parsimony phylogenetic tree problem using quantum algorithms, the combinatorial optimization model is mapped to a physical system. The model is treated as a Hamiltonian operator, where the optimal solution corresponds to the ground state of Hamiltonian \cite{mcclean2016theory}.

We employ two prominent algorithms designed to find this ground state: the Quantum Approximate Optimization Algorithm (QAOA) \cite{farhi2014quantum} and the Variational Quantum Eigensolver (VQE) \cite{peruzzo2014variational}.
To benchmark their performance, we compare their results against the exact ground state energy, which is pre-calculated via classical diagonalization. This diagonalization provides the globally optimal score for each problem, serving as a definitive target for our quantum algorithms. All simulations are conducted within the Qiskit \cite{qiskit2024} and PennyLane \cite{bergholm2018pennylane} platforms, using a noiseless statevector simulator and the gradient-free COBYLA optimizer for variational parameter updates. It must also be emphasized that all parameters are initialized randomly without prior problem-specific tuning and the utilization of statevector simulation represents an idealized environment that excludes both shot noise and hardware noise \cite{mcclean2018barren, scriva2024challenges, ejima2025probabilistic}.

\begin{figure}[htbp]
    \centering
        \includegraphics[width=0.8\linewidth]{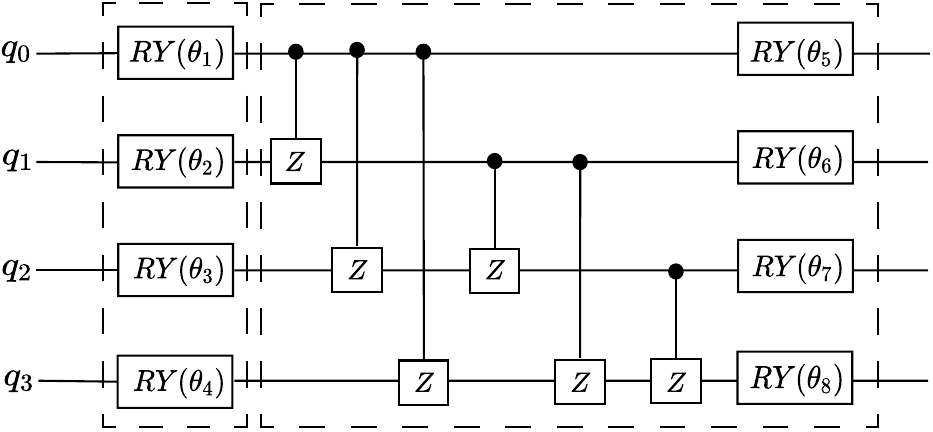}
        \caption{Structure of the hardware-efficient ansatz used in our VQE implementation, shown for $n=4$ qubits and $p=1$. It consists of layers of single-qubit Y-rotations interspersed with a layer of entangling controlled-Z gates.}
    \label{fig:hardware_efficient_ansatz}
\end{figure}

Since the performance of VQE is highly dependent on the chosen parameterized quantum circuit \cite{tilly2022variational}. We employ the widely-used hardware-efficient ansatz (HEA) \cite{kandala2017hardware}, which is constructed from alternating layers of single-qubit rotations and two-qubit entangling gates to suit near-term devices, as illustrated in Fig. \ref{fig:hardware_efficient_ansatz}.

\begin{figure}[htbp]
    \centering
        \includegraphics[width=0.8\linewidth]{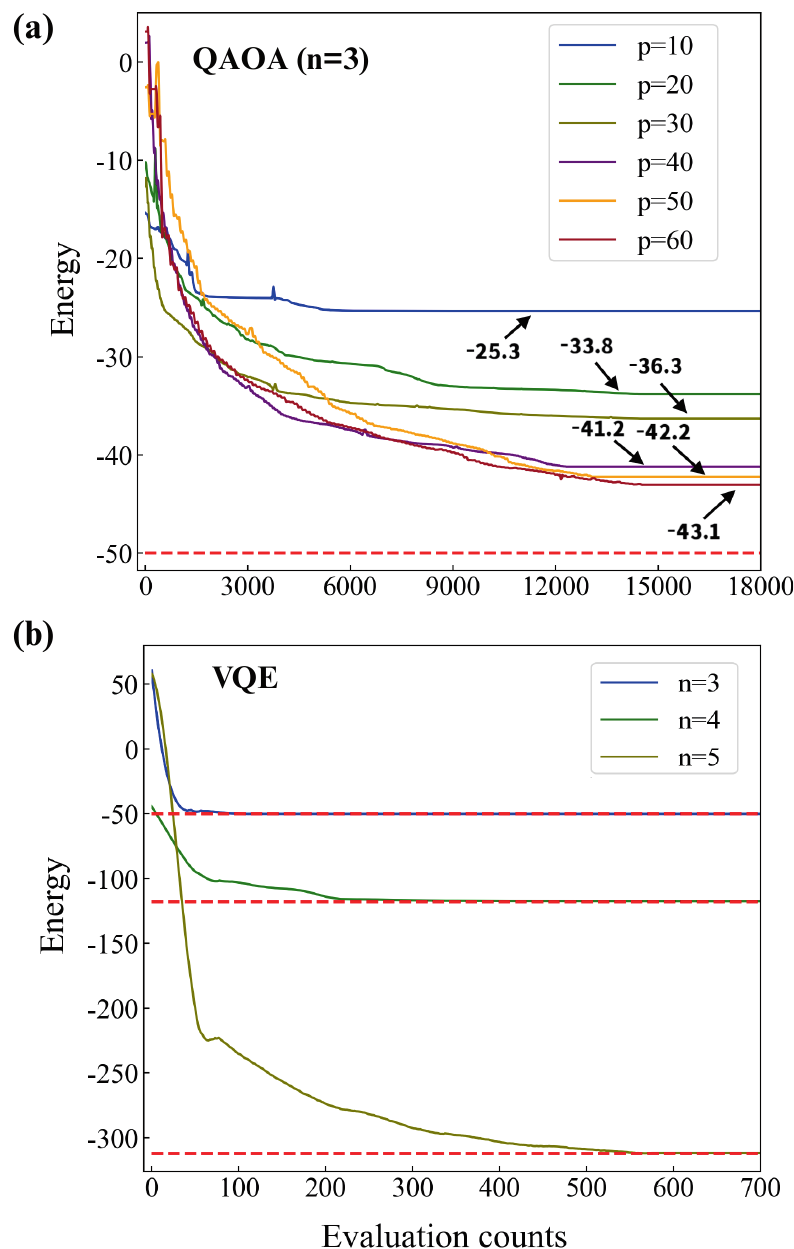}
        \caption{Comparative performance of QAOA and VQE. (a) Energy convergence of QAOA for a three leaves problem with varying circuit depths. Deeper circuits result in lower final energies but fail to reach the ground state energy (red dashed line). (b) Energy convergence of VQE for problems with different numbers of leaf nodes. The algorithm successfully finds the theoretical ground state energy for each problem size tested.}
    \label{fig:energy}
\end{figure}

The comparative performance of the VQE and QAOA algorithms is presented in Fig. \ref{fig:energy}. We first analyze the performance of QAOA. The results indicate that increasing the circuit depth allows the algorithm to converge to lower final energies.This corresponds to finding solutions with lower parsimony scores, which represent more parsimonious and biologically plausible phylogenetic tree structures. However, it consistently fails to reach the true ground state energy, evidently becoming trapped in local optima. Furthermore, increasing the circuit depth and number of parameters poses significant challenges for execution on real hardware, increasing sensitivity to noise.

In contrast, VQE paired with a HEA rapidly converged to the exact theoretical ground state energy for all tests. This performance disparity is likely attributable to the increased ruggedness of the high-order energy landscape coupled with the use of initial random parameters. To effectively traverse such rugged landscapes, the standard QAOA typically requires increased circuit depth combined with problem-informed parameter initialization \cite{zhou2020quantum, stkechly2023connecting, boy2024energy}.

While our quantum simulations are restricted to extremely small-scale instances due to current computational resource constraints, these experiments serve as a fundamental proof-of-concept, demonstrating that using variational quantum algorithms to solve the maximum parsimony phylogenetic tree problem is a feasible and effective pathway. However, extrapolating from the resource analysis in Table \ref{tab:model_comparison} and Table \ref{tab:model_complexity}, a practical demonstration on realistic biological datasets would likely necessitate hundreds of thousands of qubits. Such resource requirements are evidently beyond the reach of current NISQ devices. Nevertheless, with the continuous maturation of quantum hardware and further simplification of the mathematical model, leveraging quantum computing to address these intractable problems in real-world scenarios remains a promising avenue.

%% file: Discussion.tex
\section{Discussion}
In this paper, we design three novel combinatorial optimization models to reconstruct maximum parsimony phylogenetic trees. These models simultaneously infer ancestral sequences while searching for the tree topology, which not only circumvents the need to pre-construct internal nodes but also ensures the search space contains all possible solutions. Among them, the branch-based model through an ingenious variable definition drastically reduces the required number of variables and explicit constraints. It can define a specific tree topology with just two constraints, which is beneficial not only for solving this problem but also offers a new modeling approach for other tree problems.

The correctness of the model is validated using a high-performance classical solver on both single-site and real biological sequence fragments, confirming that the model obtains solutions of a quality superior to those from heuristics. Meanwhile, the scaling bottlenecks of classical computation motivate the exploration of new paradigms. Consequently, we map the model to a quantum Hamiltonian and solve small-scale instances with two variational quantum algorithms. In our idealized simulations, the ability of VQE paired with a HEA to rapidly find the optimal solution demonstrates the preliminary feasibility of applying quantum computing to this problem.

While VQE appears to perform better than QAOA, it is crucial to emphasize that these results rely on noiseless state-vector simulations, which neglect the effects of any noise. However, even when accounting solely for shot noise, the computational cost of variational quantum algorithms scales prohibitively with problem size, thereby limiting their scalability \cite{scriva2024challenges, barligea2025scalability}. Meanwhile, many works have demonstrated that employing improved initial states, parameter initialization methods, and parameter update strategies can significantly enhance QAOA’s convergence and solution quality \cite{zhou2020quantum, blekos2024review}.

Despite the promising results, this work has several avenues for future improvement and exploration. On the biological front, the model can be applied to phylogenomic analyses using concatenated multi-gene datasets. This approach mitigates the stochastic errors found in single-gene studies and leads to the inference of more reliable species trees \cite{wu2008simple,kapli2020phylogenetic}. Computationally,
future research will focus on enhancing the scalability and practical applicability of our method. One key direction involves exploring superior modeling approaches, ensuring compatibility with hardware restricted to two-body interactions, such as quantum annealers \cite{rajak2023quantum}. Concurrently, we will conduct rigorous performance evaluations under realistic noise conditions, specifically shot noise and hardware errors. We also plan to investigate recent algorithmic variants and advanced initialization strategies to further optimize solution quality and convergence speed.

%% file: Methods.tex
\section{Methods}
\subsection{Maximum parsimony principle}
The maximum parsimony principle identifies the optimal phylogenetic tree as the one that explains observed character differences using the fewest possible evolutionary changes \cite{fitch1971toward}. For a given multiple sequence alignment, this principle assumes each character site evolves independently. The total parsimony score for a tree topology $T$ is the sum of the minimum substitution counts required at each individual site. To compute this efficiently, sites with identical patterns are grouped, and the total score is calculated as a sum weighted by the frequency of each unique pattern:
\begin{align}
\tag{16}
    MP(T|L) = \sum_{i=1}^{k} MP(T|D_i) \times d_i,
\end{align}
where $MP(T|D_i)$ is the parsimony score for a unique site pattern $D_i$ and $d_i$ is its frequency.

The simplest approach is unweighted parsimony, which assumes all character state changes have an equal cost. Under this assumption, the minimum number of substitutions for a given tree can be computed using the Fitch algorithm \cite{fitch1971toward}. However, the assumption of equal costs is often a biological oversimplification, as substitution rates are known to vary \cite{brown1982mitochondrial,wakeley1994substitution}.

To address this limitation, weighted maximum parsimony introduces a step matrix that assigns differential costs to different types of evolutionary events, with lower costs for frequent substitutions and higher costs for rare ones \cite{nei2000molecular}. This weighting scheme not only reflects biological reality more closely but also helps mitigate systematic issues in phylogenetic reconstruction, such as long-branch attraction \cite{felsenstein1978cases}. Calculating the minimum score under such a weighted scheme requires the more general Sankoff algorithm \cite{sankoff1975minimal}, a dynamic programming approach.

\subsection{Quantum approximate optimization algorithm}
The quantum approximate optimization algorithm (QAOA) is a hybrid quantum-classical algorithm designed to find approximate solutions to combinatorial optimization problems. Inspired by quantum adiabatic evolution, QAOA provides a discretized optimization approach that is well-suited for near-term gate-based quantum computers \cite{farhi2014quantum}. The algorithm operates using two key Hamiltonians. The problem Hamiltonian ($\hat{H}_C$) encodes the classical objective function such that its ground state corresponds to the optimal solution. The mixer Hamiltonian ($\hat{H}_M$) introduces quantum fluctuations to enable exploration of the solution space.

The QAOA begins by preparing an initial state $|\psi_0^\mathrm{s}\rangle$, typically the ground state of $\hat{H}_M$. The output state $|\psi_f^\mathrm{s}\rangle$ is then prepared by alternately applying operators corresponding to $\hat{H}_C$ and $\hat{H}_M$ for $p$ layers:
\begin{equation}
\tag{17}
    |\psi_{f}^{\mathrm{s}}\rangle=\prod_{l=1}^{p}\mathrm{e}^{-i\beta_{l}\hat{H}_M}\mathrm{e}^{-i\gamma_{l}\hat{H}_C}|\psi_{0}^{\mathrm{s}}\rangle,
\end{equation}
where $(\vec{\gamma},\vec{\beta})$ are the $2p$ classical variational parameters that are optimized within a hybrid quantum-classical loop. In each iteration, the quantum computer prepares the state $|\psi_f^\mathrm{s}\rangle$ and measures its energy expectation value, $E(\vec{\gamma},\vec{\beta})=\langle\psi_f^\mathrm{s}|\hat{H}_C|\psi_f^\mathrm{s}\rangle$. This energy is then passed as a cost function to a classical optimizer, which in turn suggests an updated set of parameters designed to lower the energy. This process is iterated until the energy converges to a minimum, after which the state with the optimal parameters is prepared and measured repeatedly. The most frequently observed computational basis state is then taken as the approximate solution to the original problem.

\subsection{Variational quantum eigensolver}
The variational quantum eigensolver (VQE) is another leading hybrid quantum-classical algorithm for the noisy intermediate-scale quantum era, designed to find the lowest eigenvalue of a given Hamiltonian $\hat{H}$ \cite{peruzzo2014variational}. It is based on the Rayleigh-Ritz variational principle, which ensures that the energy expectation value of a parameterized trial state $|\psi(\vec\theta)\rangle$ provides an upper bound to the true ground-state energy $E_0$:
\begin{equation}
E(\vec{\theta}) = \langle\psi(\vec{\theta})|\hat{H}|\psi(\vec{\theta})\rangle \ge E_0.
\tag{18}\label{eq:vqe_principle}
\end{equation}
By variationally minimizing this energy, we can find a close approximation of the optimal solution.

The VQE workflow is an iterative optimization loop. Each iteration begins with a quantum processor preparing the ansatz state $|\psi(\vec\theta)\rangle$ for a given set of parameters $\vec\theta$ and measuring its energy expectation value, $E(\vec\theta)=\langle\psi(\vec\theta)|\hat{H}|\psi(\vec\theta)\rangle$. This energy is then fed as a cost function to a classical optimizer, which provides an updated set of parameters to lower the energy in the next iteration. After the loop converges to the optimal parameters $\vec\theta^*$, the final state $|\psi(\vec\theta^*)\rangle$ is prepared and measured repeatedly to identify the most probable bitstring, which corresponds to the optimal solution for a combinatorial optimization problem.

%% file: main.bbl
\begin{thebibliography}{57}
\expandafter\ifx\csname natexlab\endcsname\relax\def\natexlab#1{#1}\fi
\expandafter\ifx\csname bibnamefont\endcsname\relax
  \def\bibnamefont#1{#1}\fi
\expandafter\ifx\csname bibfnamefont\endcsname\relax
  \def\bibfnamefont#1{#1}\fi
\expandafter\ifx\csname citenamefont\endcsname\relax
  \def\citenamefont#1{#1}\fi
\expandafter\ifx\csname url\endcsname\relax
  \def\url#1{\texttt{#1}}\fi
\expandafter\ifx\csname urlprefix\endcsname\relax\def\urlprefix{URL }\fi
\providecommand{\bibinfo}[2]{#2}
\providecommand{\eprint}[2][]{\url{#2}}

\bibitem[{\citenamefont{Felsenstein}(2004)}]{felsenstein2004inferring}
\bibinfo{author}{\bibfnamefont{J.}~\bibnamefont{Felsenstein}}, in \emph{\bibinfo{booktitle}{Inferring phylogenies}} (\bibinfo{year}{2004}), pp. \bibinfo{pages}{664--664}.

\bibitem[{\citenamefont{Bull and Wichman}(2001)}]{bull2001applied}
\bibinfo{author}{\bibfnamefont{J.}~\bibnamefont{Bull}} \bibnamefont{and} \bibinfo{author}{\bibfnamefont{H.}~\bibnamefont{Wichman}}, \bibinfo{journal}{Annual Review of Ecology and systematics} \textbf{\bibinfo{volume}{32}}, \bibinfo{pages}{183} (\bibinfo{year}{2001}).

\bibitem[{\citenamefont{Davies et~al.}(2008)\citenamefont{Davies, Fritz, Grenyer, Orme, Bielby, Bininda-Emonds, Cardillo, Jones, Gittleman, Mace et~al.}}]{davies2008phylogenetic}
\bibinfo{author}{\bibfnamefont{T.~J.} \bibnamefont{Davies}}, \bibinfo{author}{\bibfnamefont{S.~A.} \bibnamefont{Fritz}}, \bibinfo{author}{\bibfnamefont{R.}~\bibnamefont{Grenyer}}, \bibinfo{author}{\bibfnamefont{C.~D.~L.} \bibnamefont{Orme}}, \bibinfo{author}{\bibfnamefont{J.}~\bibnamefont{Bielby}}, \bibinfo{author}{\bibfnamefont{O.~R.} \bibnamefont{Bininda-Emonds}}, \bibinfo{author}{\bibfnamefont{M.}~\bibnamefont{Cardillo}}, \bibinfo{author}{\bibfnamefont{K.~E.} \bibnamefont{Jones}}, \bibinfo{author}{\bibfnamefont{J.~L.} \bibnamefont{Gittleman}}, \bibinfo{author}{\bibfnamefont{G.~M.} \bibnamefont{Mace}}, \bibnamefont{et~al.}, \bibinfo{journal}{Proceedings of the National Academy of Sciences} \textbf{\bibinfo{volume}{105}}, \bibinfo{pages}{11556} (\bibinfo{year}{2008}).

\bibitem[{\citenamefont{O’Donoghue and Yordanova}(2020)}]{o2020corbomycin}
\bibinfo{author}{\bibfnamefont{N.}~\bibnamefont{O’Donoghue}} \bibnamefont{and} \bibinfo{author}{\bibfnamefont{R.}~\bibnamefont{Yordanova}}, \bibinfo{journal}{Trakia Journal of Sciences} \textbf{\bibinfo{volume}{18}}, \bibinfo{pages}{118} (\bibinfo{year}{2020}).

\bibitem[{\citenamefont{Saslis-Lagoudakis et~al.}(2011)\citenamefont{Saslis-Lagoudakis, Klitgaard, Forest, Francis, Savolainen, Williamson, and Hawkins}}]{saslis2011use}
\bibinfo{author}{\bibfnamefont{C.~H.} \bibnamefont{Saslis-Lagoudakis}}, \bibinfo{author}{\bibfnamefont{B.~B.} \bibnamefont{Klitgaard}}, \bibinfo{author}{\bibfnamefont{F.}~\bibnamefont{Forest}}, \bibinfo{author}{\bibfnamefont{L.}~\bibnamefont{Francis}}, \bibinfo{author}{\bibfnamefont{V.}~\bibnamefont{Savolainen}}, \bibinfo{author}{\bibfnamefont{E.~M.} \bibnamefont{Williamson}}, \bibnamefont{and} \bibinfo{author}{\bibfnamefont{J.~A.} \bibnamefont{Hawkins}}, \bibinfo{journal}{PloS one} \textbf{\bibinfo{volume}{6}}, \bibinfo{pages}{e22275} (\bibinfo{year}{2011}).

\bibitem[{\citenamefont{Cancino and Delbem}(2010)}]{cancino2010multi}
\bibinfo{author}{\bibfnamefont{W.}~\bibnamefont{Cancino}} \bibnamefont{and} \bibinfo{author}{\bibfnamefont{A.~C.~B.} \bibnamefont{Delbem}}, \bibinfo{journal}{New Achievements in Evolutionary Computation} pp. \bibinfo{pages}{135--156} (\bibinfo{year}{2010}).

\bibitem[{\citenamefont{Puttick et~al.}(2019)\citenamefont{Puttick, O'Reilly, Pisani, and Donoghue}}]{puttick2019probabilistic}
\bibinfo{author}{\bibfnamefont{M.~N.} \bibnamefont{Puttick}}, \bibinfo{author}{\bibfnamefont{J.~E.} \bibnamefont{O'Reilly}}, \bibinfo{author}{\bibfnamefont{D.}~\bibnamefont{Pisani}}, \bibnamefont{and} \bibinfo{author}{\bibfnamefont{P.~C.} \bibnamefont{Donoghue}}, \bibinfo{journal}{Palaeontology} \textbf{\bibinfo{volume}{62}}, \bibinfo{pages}{1} (\bibinfo{year}{2019}).

\bibitem[{\citenamefont{Day et~al.}(1986)\citenamefont{Day, Johnson, and Sankoff}}]{day1986computational}
\bibinfo{author}{\bibfnamefont{W.~H.} \bibnamefont{Day}}, \bibinfo{author}{\bibfnamefont{D.~S.} \bibnamefont{Johnson}}, \bibnamefont{and} \bibinfo{author}{\bibfnamefont{D.}~\bibnamefont{Sankoff}}, \bibinfo{journal}{Mathematical biosciences} \textbf{\bibinfo{volume}{81}}, \bibinfo{pages}{33} (\bibinfo{year}{1986}).

\bibitem[{\citenamefont{Nei and Kumar}(2000)}]{nei2000molecular}
\bibinfo{author}{\bibfnamefont{M.}~\bibnamefont{Nei}} \bibnamefont{and} \bibinfo{author}{\bibfnamefont{S.}~\bibnamefont{Kumar}}, \emph{\bibinfo{title}{Molecular evolution and phylogenetics}} (\bibinfo{publisher}{Oxford university press}, \bibinfo{year}{2000}).

\bibitem[{\citenamefont{Kirkup and Kim}(2000)}]{kirkup2000rolling}
\bibinfo{author}{\bibfnamefont{B.}~\bibnamefont{Kirkup}} \bibnamefont{and} \bibinfo{author}{\bibfnamefont{J.}~\bibnamefont{Kim}}, \bibinfo{journal}{Unpublished manuscript. Department of Ecology and Evolutionary Biology, Yale University, USA} \textbf{\bibinfo{volume}{26}} (\bibinfo{year}{2000}).

\bibitem[{\citenamefont{Emani et~al.}(2021)\citenamefont{Emani, Warrell, Anticevic, Bekiranov, Gandal, McConnell, Sapiro, Aspuru-Guzik, Baker, Bastiani et~al.}}]{emani2021quantum}
\bibinfo{author}{\bibfnamefont{P.~S.} \bibnamefont{Emani}}, \bibinfo{author}{\bibfnamefont{J.}~\bibnamefont{Warrell}}, \bibinfo{author}{\bibfnamefont{A.}~\bibnamefont{Anticevic}}, \bibinfo{author}{\bibfnamefont{S.}~\bibnamefont{Bekiranov}}, \bibinfo{author}{\bibfnamefont{M.}~\bibnamefont{Gandal}}, \bibinfo{author}{\bibfnamefont{M.~J.} \bibnamefont{McConnell}}, \bibinfo{author}{\bibfnamefont{G.}~\bibnamefont{Sapiro}}, \bibinfo{author}{\bibfnamefont{A.}~\bibnamefont{Aspuru-Guzik}}, \bibinfo{author}{\bibfnamefont{J.~T.} \bibnamefont{Baker}}, \bibinfo{author}{\bibfnamefont{M.}~\bibnamefont{Bastiani}}, \bibnamefont{et~al.}, \bibinfo{journal}{Nature Methods} \textbf{\bibinfo{volume}{18}}, \bibinfo{pages}{701} (\bibinfo{year}{2021}).

\bibitem[{\citenamefont{Fedorov and Gelfand}(2021)}]{fedorov2021towards}
\bibinfo{author}{\bibfnamefont{A.~K.} \bibnamefont{Fedorov}} \bibnamefont{and} \bibinfo{author}{\bibfnamefont{M.~S.} \bibnamefont{Gelfand}}, \bibinfo{journal}{Nature Computational Science} \textbf{\bibinfo{volume}{1}}, \bibinfo{pages}{114} (\bibinfo{year}{2021}).

\bibitem[{\citenamefont{Marchetti et~al.}(2022)\citenamefont{Marchetti, Nifos{\`\i}, Martelli, Da~Pozzo, Cappello, Banterle, Trincavelli, Martini, and D’Elia}}]{marchetti2022quantum}
\bibinfo{author}{\bibfnamefont{L.}~\bibnamefont{Marchetti}}, \bibinfo{author}{\bibfnamefont{R.}~\bibnamefont{Nifos{\`\i}}}, \bibinfo{author}{\bibfnamefont{P.~L.} \bibnamefont{Martelli}}, \bibinfo{author}{\bibfnamefont{E.}~\bibnamefont{Da~Pozzo}}, \bibinfo{author}{\bibfnamefont{V.}~\bibnamefont{Cappello}}, \bibinfo{author}{\bibfnamefont{F.}~\bibnamefont{Banterle}}, \bibinfo{author}{\bibfnamefont{M.~L.} \bibnamefont{Trincavelli}}, \bibinfo{author}{\bibfnamefont{C.}~\bibnamefont{Martini}}, \bibnamefont{and} \bibinfo{author}{\bibfnamefont{M.}~\bibnamefont{D’Elia}}, \bibinfo{journal}{Briefings in Bioinformatics} \textbf{\bibinfo{volume}{23}}, \bibinfo{pages}{bbac437} (\bibinfo{year}{2022}).

\bibitem[{\citenamefont{Na{\l}{\k{e}}cz-Charkiewicz et~al.}(2024)\citenamefont{Na{\l}{\k{e}}cz-Charkiewicz, Charkiewicz, and Nowak}}]{nalkecz2024quantum}
\bibinfo{author}{\bibfnamefont{K.}~\bibnamefont{Na{\l}{\k{e}}cz-Charkiewicz}}, \bibinfo{author}{\bibfnamefont{K.}~\bibnamefont{Charkiewicz}}, \bibnamefont{and} \bibinfo{author}{\bibfnamefont{R.~M.} \bibnamefont{Nowak}}, \bibinfo{journal}{Briefings in Bioinformatics} \textbf{\bibinfo{volume}{25}} (\bibinfo{year}{2024}).

\bibitem[{\citenamefont{Damborsky et~al.}(2025)\citenamefont{Damborsky, Kouba, Sivic, Vasina, Bednar, and Mazurenko}}]{damborsky2025quantum}
\bibinfo{author}{\bibfnamefont{J.}~\bibnamefont{Damborsky}}, \bibinfo{author}{\bibfnamefont{P.}~\bibnamefont{Kouba}}, \bibinfo{author}{\bibfnamefont{J.}~\bibnamefont{Sivic}}, \bibinfo{author}{\bibfnamefont{M.}~\bibnamefont{Vasina}}, \bibinfo{author}{\bibfnamefont{D.}~\bibnamefont{Bednar}}, \bibnamefont{and} \bibinfo{author}{\bibfnamefont{S.}~\bibnamefont{Mazurenko}}, \bibinfo{journal}{Nature Catalysis} \textbf{\bibinfo{volume}{8}}, \bibinfo{pages}{872} (\bibinfo{year}{2025}).

\bibitem[{\citenamefont{Bose et~al.}(2026)\citenamefont{Bose, Rhrissorrakrai, Utro, and Parida}}]{bose2026advancing}
\bibinfo{author}{\bibfnamefont{A.}~\bibnamefont{Bose}}, \bibinfo{author}{\bibfnamefont{K.}~\bibnamefont{Rhrissorrakrai}}, \bibinfo{author}{\bibfnamefont{F.}~\bibnamefont{Utro}}, \bibnamefont{and} \bibinfo{author}{\bibfnamefont{L.}~\bibnamefont{Parida}}, \bibinfo{journal}{Nature Reviews Molecular Cell Biology} pp. \bibinfo{pages}{1--15} (\bibinfo{year}{2026}).

\bibitem[{\citenamefont{Maurizio and Mazzola}(2025)}]{maurizio2025quantum}
\bibinfo{author}{\bibfnamefont{A.}~\bibnamefont{Maurizio}} \bibnamefont{and} \bibinfo{author}{\bibfnamefont{G.}~\bibnamefont{Mazzola}}, \bibinfo{journal}{PRX Life} \textbf{\bibinfo{volume}{3}}, \bibinfo{pages}{047001} (\bibinfo{year}{2025}).

\bibitem[{\citenamefont{Shor}(1999)}]{shor1999polynomial}
\bibinfo{author}{\bibfnamefont{P.~W.} \bibnamefont{Shor}}, \bibinfo{journal}{SIAM review} \textbf{\bibinfo{volume}{41}}, \bibinfo{pages}{303} (\bibinfo{year}{1999}).

\bibitem[{\citenamefont{Nielsen and Chuang}(2010)}]{nielsen2010quantum}
\bibinfo{author}{\bibfnamefont{M.~A.} \bibnamefont{Nielsen}} \bibnamefont{and} \bibinfo{author}{\bibfnamefont{I.~L.} \bibnamefont{Chuang}}, \emph{\bibinfo{title}{Quantum computation and quantum information}} (\bibinfo{publisher}{Cambridge university press}, \bibinfo{year}{2010}).

\bibitem[{\citenamefont{Montanaro}(2016)}]{montanaro2016quantum}
\bibinfo{author}{\bibfnamefont{A.}~\bibnamefont{Montanaro}}, \bibinfo{journal}{npj Quantum Information} \textbf{\bibinfo{volume}{2}}, \bibinfo{pages}{1} (\bibinfo{year}{2016}).

\bibitem[{\citenamefont{Rastegin}(2017)}]{rastegin2017role}
\bibinfo{author}{\bibfnamefont{A.~E.} \bibnamefont{Rastegin}}, \bibinfo{journal}{arXiv preprint arXiv:1703.10118}  (\bibinfo{year}{2017}).

\bibitem[{\citenamefont{Dalzell et~al.}(2023)\citenamefont{Dalzell, McArdle, Berta, Bienias, Chen, Gily{\'e}n, Hann, Kastoryano, Khabiboulline, Kubica et~al.}}]{dalzell2023quantum}
\bibinfo{author}{\bibfnamefont{A.~M.} \bibnamefont{Dalzell}}, \bibinfo{author}{\bibfnamefont{S.}~\bibnamefont{McArdle}}, \bibinfo{author}{\bibfnamefont{M.}~\bibnamefont{Berta}}, \bibinfo{author}{\bibfnamefont{P.}~\bibnamefont{Bienias}}, \bibinfo{author}{\bibfnamefont{C.-F.} \bibnamefont{Chen}}, \bibinfo{author}{\bibfnamefont{A.}~\bibnamefont{Gily{\'e}n}}, \bibinfo{author}{\bibfnamefont{C.~T.} \bibnamefont{Hann}}, \bibinfo{author}{\bibfnamefont{M.~J.} \bibnamefont{Kastoryano}}, \bibinfo{author}{\bibfnamefont{E.~T.} \bibnamefont{Khabiboulline}}, \bibinfo{author}{\bibfnamefont{A.}~\bibnamefont{Kubica}}, \bibnamefont{et~al.}, \bibinfo{journal}{arXiv preprint arXiv:2310.03011}  (\bibinfo{year}{2023}).

\bibitem[{\citenamefont{Preskill}(2018)}]{preskill2018quantum}
\bibinfo{author}{\bibfnamefont{J.}~\bibnamefont{Preskill}}, \bibinfo{journal}{Quantum} \textbf{\bibinfo{volume}{2}}, \bibinfo{pages}{79} (\bibinfo{year}{2018}).

\bibitem[{\citenamefont{Bharti et~al.}(2022)\citenamefont{Bharti, Cervera-Lierta, Kyaw, Haug, Alperin-Lea, Anand, Degroote, Heimonen, Kottmann, Menke et~al.}}]{bharti2022noisy}
\bibinfo{author}{\bibfnamefont{K.}~\bibnamefont{Bharti}}, \bibinfo{author}{\bibfnamefont{A.}~\bibnamefont{Cervera-Lierta}}, \bibinfo{author}{\bibfnamefont{T.~H.} \bibnamefont{Kyaw}}, \bibinfo{author}{\bibfnamefont{T.}~\bibnamefont{Haug}}, \bibinfo{author}{\bibfnamefont{S.}~\bibnamefont{Alperin-Lea}}, \bibinfo{author}{\bibfnamefont{A.}~\bibnamefont{Anand}}, \bibinfo{author}{\bibfnamefont{M.}~\bibnamefont{Degroote}}, \bibinfo{author}{\bibfnamefont{H.}~\bibnamefont{Heimonen}}, \bibinfo{author}{\bibfnamefont{J.~S.} \bibnamefont{Kottmann}}, \bibinfo{author}{\bibfnamefont{T.}~\bibnamefont{Menke}}, \bibnamefont{et~al.}, \bibinfo{journal}{Reviews of Modern Physics} \textbf{\bibinfo{volume}{94}}, \bibinfo{pages}{015004} (\bibinfo{year}{2022}).

\bibitem[{\citenamefont{Gemeinhardt et~al.}(2023)\citenamefont{Gemeinhardt, Garmendia, Wimmer, Weder, and Leymann}}]{gemeinhardt2023quantum}
\bibinfo{author}{\bibfnamefont{F.}~\bibnamefont{Gemeinhardt}}, \bibinfo{author}{\bibfnamefont{A.}~\bibnamefont{Garmendia}}, \bibinfo{author}{\bibfnamefont{M.}~\bibnamefont{Wimmer}}, \bibinfo{author}{\bibfnamefont{B.}~\bibnamefont{Weder}}, \bibnamefont{and} \bibinfo{author}{\bibfnamefont{F.}~\bibnamefont{Leymann}}, \bibinfo{journal}{ACM Computing Surveys} \textbf{\bibinfo{volume}{56}}, \bibinfo{pages}{1} (\bibinfo{year}{2023}).

\bibitem[{\citenamefont{Abbas et~al.}(2024)\citenamefont{Abbas, Ambainis, Augustino, B{\"a}rtschi, Buhrman, Coffrin, Cortiana, Dunjko, Egger, Elmegreen et~al.}}]{abbas2024challenges}
\bibinfo{author}{\bibfnamefont{A.}~\bibnamefont{Abbas}}, \bibinfo{author}{\bibfnamefont{A.}~\bibnamefont{Ambainis}}, \bibinfo{author}{\bibfnamefont{B.}~\bibnamefont{Augustino}}, \bibinfo{author}{\bibfnamefont{A.}~\bibnamefont{B{\"a}rtschi}}, \bibinfo{author}{\bibfnamefont{H.}~\bibnamefont{Buhrman}}, \bibinfo{author}{\bibfnamefont{C.}~\bibnamefont{Coffrin}}, \bibinfo{author}{\bibfnamefont{G.}~\bibnamefont{Cortiana}}, \bibinfo{author}{\bibfnamefont{V.}~\bibnamefont{Dunjko}}, \bibinfo{author}{\bibfnamefont{D.~J.} \bibnamefont{Egger}}, \bibinfo{author}{\bibfnamefont{B.~G.} \bibnamefont{Elmegreen}}, \bibnamefont{et~al.}, \bibinfo{journal}{Nature Reviews Physics} pp. \bibinfo{pages}{1--18} (\bibinfo{year}{2024}).

\bibitem[{\citenamefont{Blekos et~al.}(2024)\citenamefont{Blekos, Brand, Ceschini, Chou, Li, Pandya, and Summer}}]{blekos2024review}
\bibinfo{author}{\bibfnamefont{K.}~\bibnamefont{Blekos}}, \bibinfo{author}{\bibfnamefont{D.}~\bibnamefont{Brand}}, \bibinfo{author}{\bibfnamefont{A.}~\bibnamefont{Ceschini}}, \bibinfo{author}{\bibfnamefont{C.-H.} \bibnamefont{Chou}}, \bibinfo{author}{\bibfnamefont{R.-H.} \bibnamefont{Li}}, \bibinfo{author}{\bibfnamefont{K.}~\bibnamefont{Pandya}}, \bibnamefont{and} \bibinfo{author}{\bibfnamefont{A.}~\bibnamefont{Summer}}, \bibinfo{journal}{Physics Reports} \textbf{\bibinfo{volume}{1068}}, \bibinfo{pages}{1} (\bibinfo{year}{2024}).

\bibitem[{\citenamefont{Foulds and Graham}(1982)}]{foulds1982steiner}
\bibinfo{author}{\bibfnamefont{L.~R.} \bibnamefont{Foulds}} \bibnamefont{and} \bibinfo{author}{\bibfnamefont{R.~L.} \bibnamefont{Graham}}, \bibinfo{journal}{Advances in Applied mathematics} \textbf{\bibinfo{volume}{3}}, \bibinfo{pages}{43} (\bibinfo{year}{1982}).

\bibitem[{\citenamefont{Hwang et~al.}(1992)\citenamefont{Hwang, Richards, and Winter}}]{hwang1992steiner}
\bibinfo{author}{\bibfnamefont{F.~K.} \bibnamefont{Hwang}}, \bibinfo{author}{\bibfnamefont{D.~S.} \bibnamefont{Richards}}, \bibnamefont{and} \bibinfo{author}{\bibfnamefont{P.}~\bibnamefont{Winter}}, \bibinfo{journal}{North-Holland, Amsterdam} \textbf{\bibinfo{volume}{1}}, \bibinfo{pages}{3} (\bibinfo{year}{1992}).

\bibitem[{\citenamefont{Catanzaro et~al.}(2013)\citenamefont{Catanzaro, Ravi, and Schwartz}}]{catanzaro2013mixed}
\bibinfo{author}{\bibfnamefont{D.}~\bibnamefont{Catanzaro}}, \bibinfo{author}{\bibfnamefont{R.}~\bibnamefont{Ravi}}, \bibnamefont{and} \bibinfo{author}{\bibfnamefont{R.}~\bibnamefont{Schwartz}}, \bibinfo{journal}{Algorithms for Molecular Biology} \textbf{\bibinfo{volume}{8}}, \bibinfo{pages}{1} (\bibinfo{year}{2013}).

\bibitem[{\citenamefont{Bach et~al.}(2024)\citenamefont{Bach, Nguyen, and Dung}}]{bach2024quantum}
\bibinfo{author}{\bibfnamefont{H.~H.} \bibnamefont{Bach}}, \bibinfo{author}{\bibfnamefont{D.~K.} \bibnamefont{Nguyen}}, \bibnamefont{and} \bibinfo{author}{\bibfnamefont{N.~N.~V.} \bibnamefont{Dung}}, in \emph{\bibinfo{booktitle}{International Conference on Future Data and Security Engineering}} (\bibinfo{organization}{Springer}, \bibinfo{year}{2024}), pp. \bibinfo{pages}{158--170}.

\bibitem[{\citenamefont{Hendy and Penny}(1982)}]{hendy1982branch}
\bibinfo{author}{\bibfnamefont{M.~D.} \bibnamefont{Hendy}} \bibnamefont{and} \bibinfo{author}{\bibfnamefont{D.}~\bibnamefont{Penny}}, \bibinfo{journal}{Mathematical biosciences} \textbf{\bibinfo{volume}{59}}, \bibinfo{pages}{277} (\bibinfo{year}{1982}).

\bibitem[{\citenamefont{Kannan and Wheeler}(2012)}]{kannan2012maximum}
\bibinfo{author}{\bibfnamefont{L.}~\bibnamefont{Kannan}} \bibnamefont{and} \bibinfo{author}{\bibfnamefont{W.~C.} \bibnamefont{Wheeler}}, \bibinfo{journal}{Algorithms for molecular biology} \textbf{\bibinfo{volume}{7}}, \bibinfo{pages}{1} (\bibinfo{year}{2012}).

\bibitem[{\citenamefont{Brown et~al.}(1982)\citenamefont{Brown, Prager, Wang, and Wilson}}]{brown1982mitochondrial}
\bibinfo{author}{\bibfnamefont{W.~M.} \bibnamefont{Brown}}, \bibinfo{author}{\bibfnamefont{E.~M.} \bibnamefont{Prager}}, \bibinfo{author}{\bibfnamefont{A.}~\bibnamefont{Wang}}, \bibnamefont{and} \bibinfo{author}{\bibfnamefont{A.~C.} \bibnamefont{Wilson}}, \bibinfo{journal}{Journal of molecular evolution} \textbf{\bibinfo{volume}{18}}, \bibinfo{pages}{225} (\bibinfo{year}{1982}).

\bibitem[{\citenamefont{Wakeley}(1994)}]{wakeley1994substitution}
\bibinfo{author}{\bibfnamefont{J.}~\bibnamefont{Wakeley}}, \bibinfo{journal}{Molecular Biology and Evolution} \textbf{\bibinfo{volume}{11}}, \bibinfo{pages}{436} (\bibinfo{year}{1994}).

\bibitem[{\citenamefont{Perron and Didier}()}]{cpsatlp}
\bibinfo{author}{\bibfnamefont{L.}~\bibnamefont{Perron}} \bibnamefont{and} \bibinfo{author}{\bibfnamefont{F.}~\bibnamefont{Didier}}, \emph{\bibinfo{title}{Cp-sat}}, \urlprefix\url{https://developers.google.com/optimization/cp/cp_solver/}.

\bibitem[{\citenamefont{Tamura et~al.}(2021)\citenamefont{Tamura, Stecher, and Kumar}}]{tamura2021mega11}
\bibinfo{author}{\bibfnamefont{K.}~\bibnamefont{Tamura}}, \bibinfo{author}{\bibfnamefont{G.}~\bibnamefont{Stecher}}, \bibnamefont{and} \bibinfo{author}{\bibfnamefont{S.}~\bibnamefont{Kumar}}, \bibinfo{journal}{Molecular biology and evolution} \textbf{\bibinfo{volume}{38}}, \bibinfo{pages}{3022} (\bibinfo{year}{2021}).

\bibitem[{\citenamefont{McClean et~al.}(2016)\citenamefont{McClean, Romero, Babbush, and Aspuru-Guzik}}]{mcclean2016theory}
\bibinfo{author}{\bibfnamefont{J.~R.} \bibnamefont{McClean}}, \bibinfo{author}{\bibfnamefont{J.}~\bibnamefont{Romero}}, \bibinfo{author}{\bibfnamefont{R.}~\bibnamefont{Babbush}}, \bibnamefont{and} \bibinfo{author}{\bibfnamefont{A.}~\bibnamefont{Aspuru-Guzik}}, \bibinfo{journal}{New Journal of Physics} \textbf{\bibinfo{volume}{18}}, \bibinfo{pages}{023023} (\bibinfo{year}{2016}).

\bibitem[{\citenamefont{Farhi et~al.}(2014)\citenamefont{Farhi, Goldstone, and Gutmann}}]{farhi2014quantum}
\bibinfo{author}{\bibfnamefont{E.}~\bibnamefont{Farhi}}, \bibinfo{author}{\bibfnamefont{J.}~\bibnamefont{Goldstone}}, \bibnamefont{and} \bibinfo{author}{\bibfnamefont{S.}~\bibnamefont{Gutmann}}, \bibinfo{journal}{arXiv preprint arXiv:1411.4028}  (\bibinfo{year}{2014}).

\bibitem[{\citenamefont{Peruzzo et~al.}(2014)\citenamefont{Peruzzo, McClean, Shadbolt, Yung, Zhou, Love, Aspuru-Guzik, and O’brien}}]{peruzzo2014variational}
\bibinfo{author}{\bibfnamefont{A.}~\bibnamefont{Peruzzo}}, \bibinfo{author}{\bibfnamefont{J.}~\bibnamefont{McClean}}, \bibinfo{author}{\bibfnamefont{P.}~\bibnamefont{Shadbolt}}, \bibinfo{author}{\bibfnamefont{M.-H.} \bibnamefont{Yung}}, \bibinfo{author}{\bibfnamefont{X.-Q.} \bibnamefont{Zhou}}, \bibinfo{author}{\bibfnamefont{P.~J.} \bibnamefont{Love}}, \bibinfo{author}{\bibfnamefont{A.}~\bibnamefont{Aspuru-Guzik}}, \bibnamefont{and} \bibinfo{author}{\bibfnamefont{J.~L.} \bibnamefont{O’brien}}, \bibinfo{journal}{Nature communications} \textbf{\bibinfo{volume}{5}}, \bibinfo{pages}{4213} (\bibinfo{year}{2014}).

\bibitem[{\citenamefont{Javadi-Abhari et~al.}(2024)\citenamefont{Javadi-Abhari, Treinish, Krsulich, Wood, Lishman, Gacon, Martiel, Nation, Bishop, Cross et~al.}}]{qiskit2024}
\bibinfo{author}{\bibfnamefont{A.}~\bibnamefont{Javadi-Abhari}}, \bibinfo{author}{\bibfnamefont{M.}~\bibnamefont{Treinish}}, \bibinfo{author}{\bibfnamefont{K.}~\bibnamefont{Krsulich}}, \bibinfo{author}{\bibfnamefont{C.~J.} \bibnamefont{Wood}}, \bibinfo{author}{\bibfnamefont{J.}~\bibnamefont{Lishman}}, \bibinfo{author}{\bibfnamefont{J.}~\bibnamefont{Gacon}}, \bibinfo{author}{\bibfnamefont{S.}~\bibnamefont{Martiel}}, \bibinfo{author}{\bibfnamefont{P.~D.} \bibnamefont{Nation}}, \bibinfo{author}{\bibfnamefont{L.~S.} \bibnamefont{Bishop}}, \bibinfo{author}{\bibfnamefont{A.~W.} \bibnamefont{Cross}}, \bibnamefont{et~al.}, \emph{\bibinfo{title}{Quantum computing with {Q}iskit}} (\bibinfo{year}{2024}), \eprint{2405.08810}.

\bibitem[{\citenamefont{Bergholm et~al.}(2018)\citenamefont{Bergholm, Izaac, Schuld, Gogolin, Ahmed, Ajith, Alam, Alonso-Linaje, AkashNarayanan, Asadi et~al.}}]{bergholm2018pennylane}
\bibinfo{author}{\bibfnamefont{V.}~\bibnamefont{Bergholm}}, \bibinfo{author}{\bibfnamefont{J.}~\bibnamefont{Izaac}}, \bibinfo{author}{\bibfnamefont{M.}~\bibnamefont{Schuld}}, \bibinfo{author}{\bibfnamefont{C.}~\bibnamefont{Gogolin}}, \bibinfo{author}{\bibfnamefont{S.}~\bibnamefont{Ahmed}}, \bibinfo{author}{\bibfnamefont{V.}~\bibnamefont{Ajith}}, \bibinfo{author}{\bibfnamefont{M.~S.} \bibnamefont{Alam}}, \bibinfo{author}{\bibfnamefont{G.}~\bibnamefont{Alonso-Linaje}}, \bibinfo{author}{\bibfnamefont{B.}~\bibnamefont{AkashNarayanan}}, \bibinfo{author}{\bibfnamefont{A.}~\bibnamefont{Asadi}}, \bibnamefont{et~al.}, \bibinfo{journal}{arXiv preprint arXiv:1811.04968}  (\bibinfo{year}{2018}).

\bibitem[{\citenamefont{McClean et~al.}(2018)\citenamefont{McClean, Boixo, Smelyanskiy, Babbush, and Neven}}]{mcclean2018barren}
\bibinfo{author}{\bibfnamefont{J.~R.} \bibnamefont{McClean}}, \bibinfo{author}{\bibfnamefont{S.}~\bibnamefont{Boixo}}, \bibinfo{author}{\bibfnamefont{V.~N.} \bibnamefont{Smelyanskiy}}, \bibinfo{author}{\bibfnamefont{R.}~\bibnamefont{Babbush}}, \bibnamefont{and} \bibinfo{author}{\bibfnamefont{H.}~\bibnamefont{Neven}}, \bibinfo{journal}{Nature communications} \textbf{\bibinfo{volume}{9}}, \bibinfo{pages}{4812} (\bibinfo{year}{2018}).

\bibitem[{\citenamefont{Scriva et~al.}(2024)\citenamefont{Scriva, Astrakhantsev, Pilati, and Mazzola}}]{scriva2024challenges}
\bibinfo{author}{\bibfnamefont{G.}~\bibnamefont{Scriva}}, \bibinfo{author}{\bibfnamefont{N.}~\bibnamefont{Astrakhantsev}}, \bibinfo{author}{\bibfnamefont{S.}~\bibnamefont{Pilati}}, \bibnamefont{and} \bibinfo{author}{\bibfnamefont{G.}~\bibnamefont{Mazzola}}, \bibinfo{journal}{Physical Review A} \textbf{\bibinfo{volume}{109}}, \bibinfo{pages}{032408} (\bibinfo{year}{2024}).

\bibitem[{\citenamefont{Ejima et~al.}(2025)\citenamefont{Ejima, Seki, Fauseweh, and Yunoki}}]{ejima2025probabilistic}
\bibinfo{author}{\bibfnamefont{S.}~\bibnamefont{Ejima}}, \bibinfo{author}{\bibfnamefont{K.}~\bibnamefont{Seki}}, \bibinfo{author}{\bibfnamefont{B.}~\bibnamefont{Fauseweh}}, \bibnamefont{and} \bibinfo{author}{\bibfnamefont{S.}~\bibnamefont{Yunoki}}, \bibinfo{journal}{Physical Review Research} \textbf{\bibinfo{volume}{7}}, \bibinfo{pages}{043182} (\bibinfo{year}{2025}).

\bibitem[{\citenamefont{Tilly et~al.}(2022)\citenamefont{Tilly, Chen, Cao, Picozzi, Setia, Li, Grant, Wossnig, Rungger, Booth et~al.}}]{tilly2022variational}
\bibinfo{author}{\bibfnamefont{J.}~\bibnamefont{Tilly}}, \bibinfo{author}{\bibfnamefont{H.}~\bibnamefont{Chen}}, \bibinfo{author}{\bibfnamefont{S.}~\bibnamefont{Cao}}, \bibinfo{author}{\bibfnamefont{D.}~\bibnamefont{Picozzi}}, \bibinfo{author}{\bibfnamefont{K.}~\bibnamefont{Setia}}, \bibinfo{author}{\bibfnamefont{Y.}~\bibnamefont{Li}}, \bibinfo{author}{\bibfnamefont{E.}~\bibnamefont{Grant}}, \bibinfo{author}{\bibfnamefont{L.}~\bibnamefont{Wossnig}}, \bibinfo{author}{\bibfnamefont{I.}~\bibnamefont{Rungger}}, \bibinfo{author}{\bibfnamefont{G.~H.} \bibnamefont{Booth}}, \bibnamefont{et~al.}, \bibinfo{journal}{Physics Reports} \textbf{\bibinfo{volume}{986}}, \bibinfo{pages}{1} (\bibinfo{year}{2022}).

\bibitem[{\citenamefont{Kandala et~al.}(2017)\citenamefont{Kandala, Mezzacapo, Temme, Takita, Brink, Chow, and Gambetta}}]{kandala2017hardware}
\bibinfo{author}{\bibfnamefont{A.}~\bibnamefont{Kandala}}, \bibinfo{author}{\bibfnamefont{A.}~\bibnamefont{Mezzacapo}}, \bibinfo{author}{\bibfnamefont{K.}~\bibnamefont{Temme}}, \bibinfo{author}{\bibfnamefont{M.}~\bibnamefont{Takita}}, \bibinfo{author}{\bibfnamefont{M.}~\bibnamefont{Brink}}, \bibinfo{author}{\bibfnamefont{J.~M.} \bibnamefont{Chow}}, \bibnamefont{and} \bibinfo{author}{\bibfnamefont{J.~M.} \bibnamefont{Gambetta}}, \bibinfo{journal}{nature} \textbf{\bibinfo{volume}{549}}, \bibinfo{pages}{242} (\bibinfo{year}{2017}).

\bibitem[{\citenamefont{Zhou et~al.}(2020)\citenamefont{Zhou, Wang, Choi, Pichler, and Lukin}}]{zhou2020quantum}
\bibinfo{author}{\bibfnamefont{L.}~\bibnamefont{Zhou}}, \bibinfo{author}{\bibfnamefont{S.-T.} \bibnamefont{Wang}}, \bibinfo{author}{\bibfnamefont{S.}~\bibnamefont{Choi}}, \bibinfo{author}{\bibfnamefont{H.}~\bibnamefont{Pichler}}, \bibnamefont{and} \bibinfo{author}{\bibfnamefont{M.~D.} \bibnamefont{Lukin}}, \bibinfo{journal}{Physical Review X} \textbf{\bibinfo{volume}{10}}, \bibinfo{pages}{021067} (\bibinfo{year}{2020}).

\bibitem[{\citenamefont{St{\k{e}}ch{\l}y et~al.}(2023)\citenamefont{St{\k{e}}ch{\l}y, Gao, Yogendran, Fontana, and Rudolph}}]{stkechly2023connecting}
\bibinfo{author}{\bibfnamefont{M.}~\bibnamefont{St{\k{e}}ch{\l}y}}, \bibinfo{author}{\bibfnamefont{L.}~\bibnamefont{Gao}}, \bibinfo{author}{\bibfnamefont{B.}~\bibnamefont{Yogendran}}, \bibinfo{author}{\bibfnamefont{E.}~\bibnamefont{Fontana}}, \bibnamefont{and} \bibinfo{author}{\bibfnamefont{M.}~\bibnamefont{Rudolph}}, \bibinfo{journal}{arXiv preprint arXiv:2305.13594}  (\bibinfo{year}{2023}).

\bibitem[{\citenamefont{Boy and Wales}(2024)}]{boy2024energy}
\bibinfo{author}{\bibfnamefont{C.}~\bibnamefont{Boy}} \bibnamefont{and} \bibinfo{author}{\bibfnamefont{D.~J.} \bibnamefont{Wales}}, \bibinfo{journal}{Physical Review A} \textbf{\bibinfo{volume}{109}}, \bibinfo{pages}{062602} (\bibinfo{year}{2024}).

\bibitem[{\citenamefont{B{\"a}rligea et~al.}(2025)\citenamefont{B{\"a}rligea, Poggel, and Lorenz}}]{barligea2025scalability}
\bibinfo{author}{\bibfnamefont{A.}~\bibnamefont{B{\"a}rligea}}, \bibinfo{author}{\bibfnamefont{B.}~\bibnamefont{Poggel}}, \bibnamefont{and} \bibinfo{author}{\bibfnamefont{J.~M.} \bibnamefont{Lorenz}}, \bibinfo{journal}{Physical Review A} \textbf{\bibinfo{volume}{112}}, \bibinfo{pages}{032407} (\bibinfo{year}{2025}).

\bibitem[{\citenamefont{Wu and Eisen}(2008)}]{wu2008simple}
\bibinfo{author}{\bibfnamefont{M.}~\bibnamefont{Wu}} \bibnamefont{and} \bibinfo{author}{\bibfnamefont{J.~A.} \bibnamefont{Eisen}}, \bibinfo{journal}{Genome biology} \textbf{\bibinfo{volume}{9}}, \bibinfo{pages}{1} (\bibinfo{year}{2008}).

\bibitem[{\citenamefont{Kapli et~al.}(2020)\citenamefont{Kapli, Yang, and Telford}}]{kapli2020phylogenetic}
\bibinfo{author}{\bibfnamefont{P.}~\bibnamefont{Kapli}}, \bibinfo{author}{\bibfnamefont{Z.}~\bibnamefont{Yang}}, \bibnamefont{and} \bibinfo{author}{\bibfnamefont{M.~J.} \bibnamefont{Telford}}, \bibinfo{journal}{Nature Reviews Genetics} \textbf{\bibinfo{volume}{21}}, \bibinfo{pages}{428} (\bibinfo{year}{2020}).

\bibitem[{\citenamefont{Rajak et~al.}(2023)\citenamefont{Rajak, Suzuki, Dutta, and Chakrabarti}}]{rajak2023quantum}
\bibinfo{author}{\bibfnamefont{A.}~\bibnamefont{Rajak}}, \bibinfo{author}{\bibfnamefont{S.}~\bibnamefont{Suzuki}}, \bibinfo{author}{\bibfnamefont{A.}~\bibnamefont{Dutta}}, \bibnamefont{and} \bibinfo{author}{\bibfnamefont{B.~K.} \bibnamefont{Chakrabarti}}, \bibinfo{journal}{Philosophical Transactions of the Royal Society A} \textbf{\bibinfo{volume}{381}}, \bibinfo{pages}{20210417} (\bibinfo{year}{2023}).

\bibitem[{\citenamefont{Fitch}(1971)}]{fitch1971toward}
\bibinfo{author}{\bibfnamefont{W.~M.} \bibnamefont{Fitch}}, \bibinfo{journal}{Systematic Biology} \textbf{\bibinfo{volume}{20}}, \bibinfo{pages}{406} (\bibinfo{year}{1971}).

\bibitem[{\citenamefont{Felsenstein}(1978)}]{felsenstein1978cases}
\bibinfo{author}{\bibfnamefont{J.}~\bibnamefont{Felsenstein}}, \bibinfo{journal}{Systematic zoology} \textbf{\bibinfo{volume}{27}}, \bibinfo{pages}{401} (\bibinfo{year}{1978}).

\bibitem[{\citenamefont{Sankoff}(1975)}]{sankoff1975minimal}
\bibinfo{author}{\bibfnamefont{D.}~\bibnamefont{Sankoff}}, \bibinfo{journal}{SIAM Journal on Applied Mathematics} \textbf{\bibinfo{volume}{28}}, \bibinfo{pages}{35} (\bibinfo{year}{1975}).

\end{thebibliography}
